\documentclass[aip,amsmath,amssymb,preprint]{revtex4-1}
\usepackage{graphicx}
\usepackage{dcolumn}
\usepackage{bm}

\begin{document}


\title{Interface induced perpendicular magnetic anisotropy in Co/CoO/Co thin film structure: An $\emph{in-situ}$ MOKE investigation }


\author{Dileep Kumar}
\email{dkumar@csr.res.in}
\author{Ajay Gupta}
\affiliation{ UGC-DAE Consortium for Scientific Research, Khandwa
Road, Indore-452017, India}
\author{P. Patidar}
\affiliation{School of Nanotechnology, RGPV, Bhopal-462021}
\author{K.K. Pandey}
\author{T. Sant}
\author{S.M Sharma}
\affiliation{High Pressure and synchrotron radiation physics
division, BARC, Trombay, Mumbai- 400 085}

\date{\today}

\begin{abstract}
Co /CoO/Co polycrystalline film was grown on Si (001) substrate
and magnetic properties have been investigated using
\emph{in-situ} magneto-optic Kerr effect during growth of the
sample. Magnetic anisotropy with easy axis perpendicular to the
film surface has been observed in top Co layer, whereas bottom
layer was found to be soft with in-plane magnetization without any
influence of top layer. Ex-situ in-plane and out-of-plane
diffraction measurements revealed that the growth of Co on
oxidized interface takes place with preferential orientation of
c-axis perpendicular to the film plane, which results in the
observed perpendicular magnetic anisotropy. Texturing of the
c-axis is expected to be a result of minimization of the interface
energy due to hybridization between Co and oxygen at the
interface.
\end{abstract}

\maketitle

Perpendicular magnetic anisotropy (PMA) is a newly emergent area
of thin film magnetism in which the preferential alignment of the
spins perpendicular to the film surface plays an important role in
the functioning of the spintronic devices. In the past, magnetic
materials which exhibit PMA drew wide attention because of their
use in perpendicular recording media.~\cite{Cheg:JMS:1991,
Nature06:PRM} Recently, thin film nanostructures which exhibit
strong PMA rapidly actuated extensive research efforts because of
the great current interest in perpendicular spin valves (p-SVs)
and magnetic tunnel junction (p-MTJs) devices.~\cite{Nature10}
Strong PMA appears in periodically altered ferromagnetic (FM)/
noble metal(NM) multilayers  such as Co/Pt, Co/Pd, Co/Au and
CoFeB/Pd.~\cite{PRL98:CoPt,JAP11:CoPd,PRB94:COPt-Pd-Ni,
PRL88:CoAu} In these multilayer systems, increase in the orbital
momentum of Co due to the strong hybridization between the 3d
orbitals of the transition metal and the 5d orbitals of heavy
metals at the interface is known to be responsible for the strong
PMA.~\cite{PRL98:CoPt,PRB94:COPt-Pd-Ni} As the hybridization is
localized at the interfaces, therefore strong PMA appears only
with very thin magnetic layer (typically in the range of 3 {\AA}
to 6 {\AA}  thick). Most of the PMA systems have already been used
as magnetic electrodes in magnetic tunnel junctions, however
because of limited thermal stability and ultra-thin magnetic
layer, which substantially reduces the effective spin polarization
of the electrode and leads to quite small tunnelling magneto-
resistance (TMR) value, none of them are useful from application
point of view.

Recently, L.E Nistor \emph{et al.}~\cite{nistor:APL9} studied
Pt/Co/Oxide and Oxide/Co/Pt thin film structures and obtained
strong PMA with Co layer as thick as 30 {\AA} after annealing at
350$^{0}$C. They have demonstrated that despite the degraded
growth of Co on Oxide, the Oxide/Co interface brings much more PMA
as compare to Pt/Co. The origin of enhanced value of PMA was
attributed to the alignments of the Co-O bonds due to
hybridization between Co and O orbitals at the interface. In
another study by Qin-Li\emph{et al.}~\cite{APE10:Qin-LiLv} a Co/
Pt bilayers with 15 {\AA} thick Co have been prepared by
incorporating some oxygen atoms at interface. A significant
increase in PMA after annealing at 250$^{0}$C-400$^{0}$C was
attributed to the formation of large number of Co-O-Pt or Pt-Co-O
bonds orientated normal to the interface. In this study, it has
also been demonstrated that an additional 20 {\AA} thick Co on
Co/CoO/Pt structure aligned perpendicular to the film plane due to
the exchange coupling with underneath multilayer. In fact the role
of oxygen to improve PMA opened up another avenue to realizing
thick, thermally stable electrodes for both p-SVs and p-MTJs and
therefore being studied in detail both
experimentally~\cite{PRB09:Co-Alox} and
theoretically.~\cite{PRB11:FE-Mgo} Although these results are
quite encouraging but still there is a lack of clear understanding
in this area. In this letter, we have studied the growth of thick
Co layer($~\sim$100{\AA})on uniformly oxidized CoO layer which has
been prepared by thermally oxidizing thick Co layer on Si (001)
substrate. Magnetic properties during each step of growth of the
films have been investigated using \emph{in-situ} magneto optical
Kerr effect (MOKE). In contrast to the previous studies, We have
obtained significant PMA in much higher thickness of Co film even
in as prepared stage (without post thermal annealing the
structure). Using depth resolved in-situ MOKE, it is also
demonstrated that oxide layer (CoO) of 26 {\AA} thickness nicely
separate different magnetic state of adjacent (bottom and top) two
magnetic layers.

Co/CoO/Co multilayer structure has been grown in an ultra-high
vacuum chamber in the base pressure $\sim2\times10^{-8}$ mbar with
facility for e-beam evaporation, and MOKE
measurement.~\cite{Chamber:DK}  Film thickness was measured
in-situ using a calibrated quartz crystal thickness monitor.
Growth of the sample was done on silicon substrate in three steps;
i) deposition of $\sim$500 {\AA} thick Co film at room temperature
(denoted as F1), ii) in-situ annealing at 450$^{0}$C for 1 hour in
a vacuum of $\sim5\times10^{-4}$ mbar (denoted as F2) in order to
form a thin oxide layer at the surface. Annealing is also expected
to results in transformation of hcp phase of Co to fcc
phase~\cite{JDAP09:Co-Ann}, and iii) deposition of
another$~\sim$100 {\AA} thick Co film (denoted as F3) on top of
it. In-situ hysteresis loops were measured using MOKE in
longitudinal geometry (L-MOKE) after each step of deposition (F1,
F2 and F3).  It may be noted that after the second stage, a part
of the sample (F2- film) was preserved by masking. Another Co film
on Si substrate was deposited separately in the identical
conditions, which is expected to be identical to film F1. These
films were used for ex-situ measurements.

\begin{figure}
\begin{center}
\includegraphics[angle=0, width=0.75\textwidth]{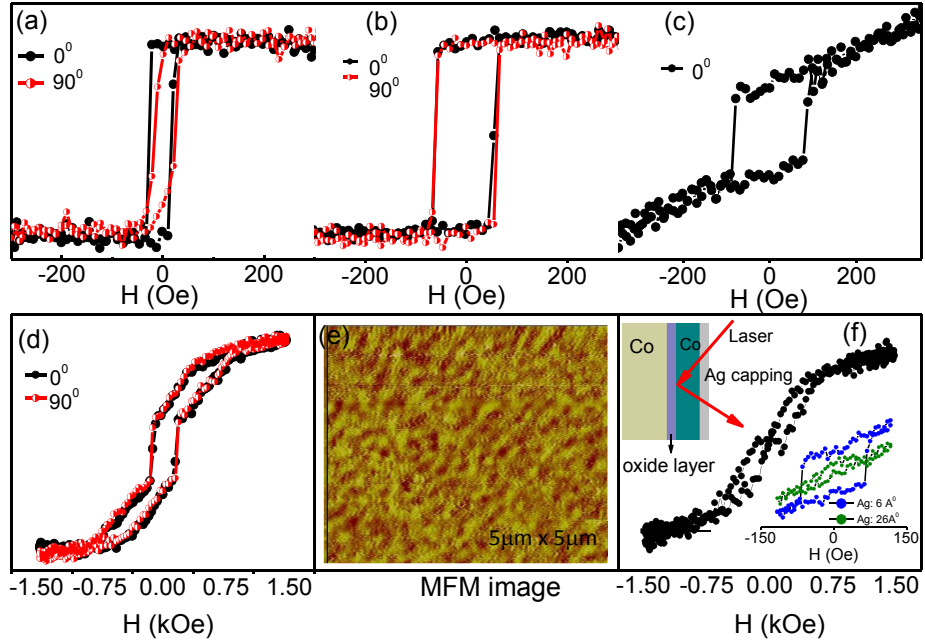}
\end{center}
\vspace {-.5 cm} \caption{\label{Fig1} (Colored online) Hysteresis
loops after the different stages of the sample growth; a) F1 b) F2
and c) F3. (d) Hysteresis loop of F3-film recorded with higher
magnetic field using ex-situ MOKE; =0$^{0}$, 90$^{0}$ angles
corresponds to hysteresis loop along easy direction and hard
direction of magnetization. (e) MFM image of the F3 film (f) loop
after 26{\AA} thick Ag layer on top of F3 film. In set of the
figure shows two representative loops, recorded in-situ with
6{\AA} and 26{\AA} thick capping layer}
\end{figure}

Figure 1 gives hysteresis loops recorded in-situ after different
stages F1, F2, and F3 of sample growth. Film F1 exhibits a square
loop with a coercive field (Hc) of 21${\pm}$0.5 Oe.  As shown in
fig. 1(a), small variation in shape with azimuthal direction was
observed, suggesting presence of a very weak in-plane uniaxial
magnetic anisotropy in F1 film.  In film F2 (fig. 1b), coercivity
increased almost 3 times as compare to F1 film and similar loops
with azimuthal direction in this figure, suggesting disappearance
of weak in-plain magnetic anisotropy, which might have induced due
to the stresses produced during the growth of the film F1. It is
important to note that, in film F3 (fig.1c), when an additional Co
layer is deposited on F2 film, coercivity remains almost the same,
but it becomes difficult to saturate the hysteresis loop using a
magnetic field of $\sim$300 Oe, which is the maximum field which
could be applied in our in-situ set-up. Hysteresis loop of the
F3-film was also measured ex-situ under higher applied magnetic
field and is given in fig.1(d). It may be noted that the
hysteresis loop is still not getting saturated upto 1500Oe. No
variation in the hysteresis loops was observed as a function of
azimuthal angle. This indicates presence of a strong perpendicular
magnetic anisotropy in F3 film. Magnetic force microscopy (MFM)
image in figure 1(e), with dense stripe magnetic domain structure,
strongly confirms the perpendicular magnetization in this film.

It is necessary to mention here that He-Ne laser of wavelength
=6328{\AA} is used for MOKE measurements, which penetrates
typically upto 200 to 300 {\AA} in depth of Co
film.~\cite{ASC-Co:Dk} Therefore, double hysteresis loop of F3
film may contain combined hysteresis loop of both bottom and top
Co layers. In order to confirm this, a small piece of F3 film was
reloaded into the chamber and an Ag film was deposited on it.
Hysteresis loops were recorded simultaneously as a function of the
thickness of nonmagnetic Ag capping layer. This provides
hysteresis loops from decreasing depth of the film because of the
laser penetration into the Co film is limited by the top Ag layer.
These measurements reveal that the soft loop in middle gradually
disappears and hard loop remains in place after ~26{\AA} thick Ag
layer on top. Hysteresis loop with 26{\AA} thick capping layer was
also recorded using ex-situ MOKE with higher magnetic field and is
shown in figure 1(f). Decreased contribution of the soft (center
loop) part with increasing thickness of Ag reveals that the bottom
Co remains soft with in-plane magnetic moments whereas the top Co
layer, which was deposited on F2 film, possesses perpendicular
magnetic anisotropy.

\begin{figure}
\begin{center}
\includegraphics[angle=0, width=0.5\textwidth]{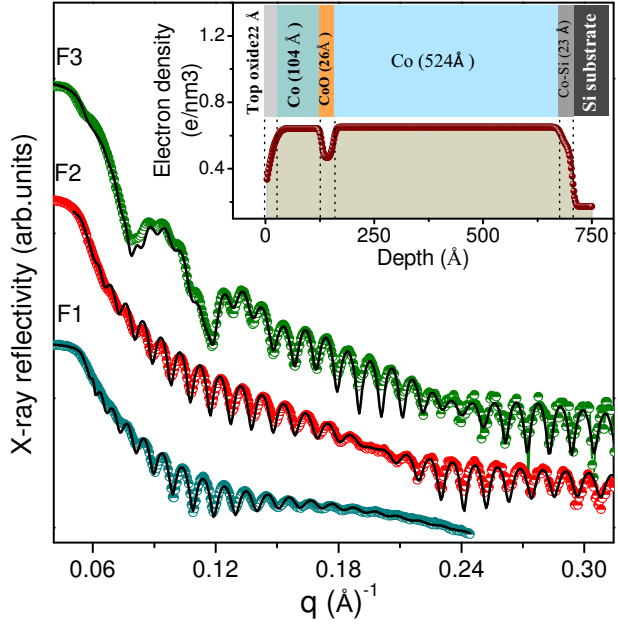}
\end{center}
\vspace {-.5 cm} \caption{\label{Fig2} (Colored online), i)
Experimental (doted symbol) and fitted (line) x-ray reflectivity
patterns of F1, F2 and F3 films. Inset gives electron density
($\rho$) profile and final structure of the sample F3 film. In
order to elucidate the origin of the observed magnetic properties
and to correlate the same with structure of the sample, x-ray
reflectivity and x-ray diffraction measurements were done on these
films}
\end{figure}

Figure 2 gives x-ray reflectivity patterns together with best fit
to the data using Parratt's formalism.~\cite{Parratt:PRB} Electron
density depth profile of sample F3 as shown in inset of the figure
2, strongly confirms the formation of 26{\AA} thick oxide layer
with electron density 28$\%$ less than that of bulk Co. In fact,
this sample also displays exchange bias, after being field cooled
to 10K, supporting the formation of native antiferromagnetic oxide
layer of CoO. It may be noted that the interfaces at both side of
oxide layer are found relatively sharp
($\sigma_\mathrm{Co-on-CoO}$ = 4.1{\AA} and
$\sigma_\mathrm{CoO-on-Co}$ = 5.2{\AA}) and are comparable to the
substrate roughness, which indicates a homogeneous thermal
oxidation of the bottom Co layer.~\cite{nistor:APL9}

Recently H.X tang \emph{et al.}~\cite{PRB11:FE-Mgo} investigated
origin of PMA at the interface between ferromagnetic transition
metals and metallic oxides via first principles theoretical
calculation. Origin of strong PMA due to the interfacial oxide in
Fe/ MgO(100) structure was attributed to the hybridization between
Fe-3d and O-2p orbitals. Experimentally, in the Pt/Co/AlOx
trilayer system, it has been observed that a substantial amount of
oxygen at the interface induced strong PMA in this system due to
Co and O atomic hybridization at the
interface.~\cite{PRB09:Co-Alox} Hybridization results Co-O
orbitals perpendicular to the interface due to minimization of
Co-O binding energy and leads to a strong PMA in this trilayer
system. A similar effect has also been observed in other systems
such as CoO/[Co/Pt], Co/native oxide/Pt and Pt/Co/Oxide
etc.~\cite{nistor:APL9,APE10:Qin-LiLv} It is important to note
that hybridization is localized at the interfaces and therefore,
even with oxygen mediated interfaces (as mentioned above),
significant PMA is observed only in ultra thin magnetic layer
ranging from few {\AA} to a few tens of {\AA}.~\cite{nistor:APL9,
APE10:Qin-LiLv} In the present work, PMA in 104{\AA} thick Co
layer is some what surprising and can not be understood only
in-terms of interface hybridization.

\begin{figure}
\begin{center}
\includegraphics[angle=0, width=0.65\textwidth]{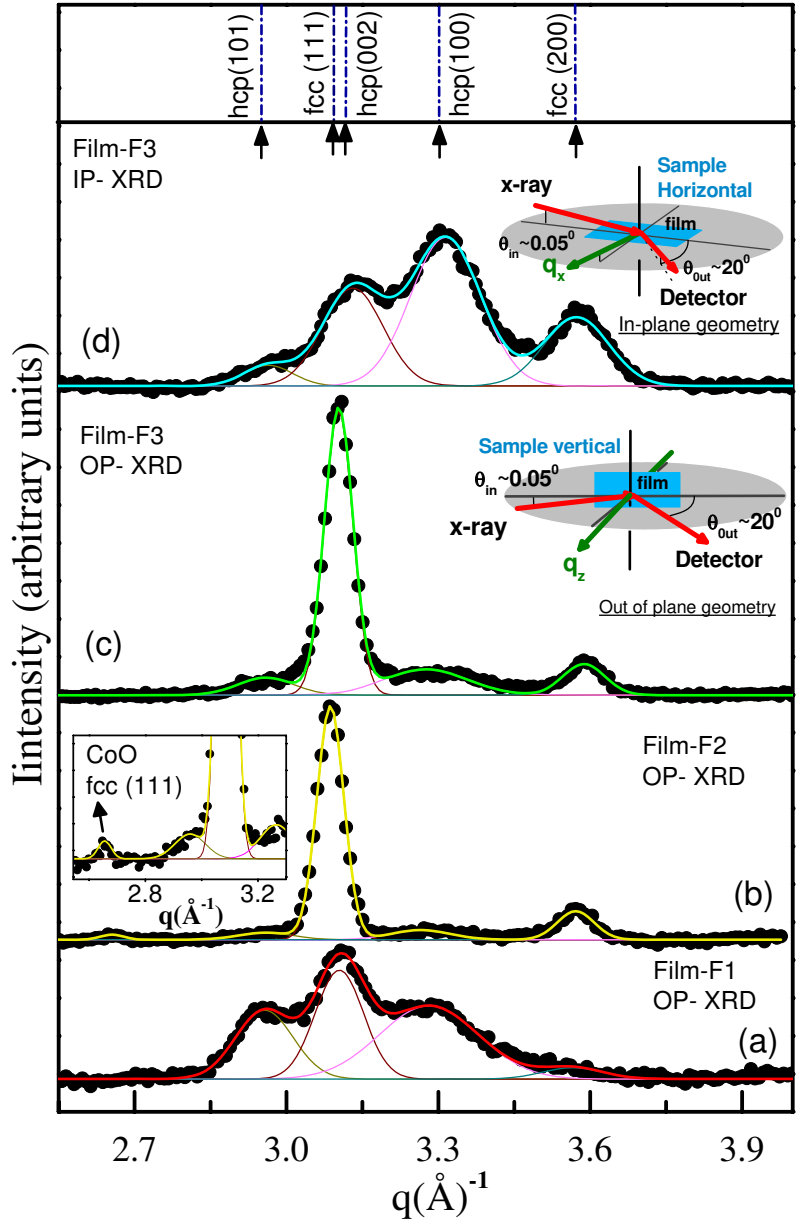}
\end{center}
\vspace {-.05 cm} \caption{\label{Fig3} (Colored online)Fitted XRD
patterns in out of plane geometry (OP-XRD) a) for F1, b) for F2
and c) for F3 sample. d) fitted in-plane XRD (IP-XRD) for F3
sample.  All samples measured at 0.05$^{o}$ incident angle.
Geometries used for the measurements are given in in-set of the
figure}
\end{figure}

In order to explore the origin of the PMA in top Co layer,
structure of the films was investigated using energy dispersive
x-ray diffraction (EDXRD) at BM-11 beamline of Indus-2,
synchrotron radiation source, Indore which is a white x-ray
beamline.~\cite{pandeyji:EDXRD} In order to find in-plane and
out-of-plane structure of sample F3, measurements were carried out
in two geometries as shown in inset of the fig.3; i) grazing
incidence out of plane (OP-XRD) geometry, where the incident beam
falls at very low grazing angle ($\theta$$_{in}$ $\sim$0.05$^{o}$
) on the sample which was kept vertically, while the detector was
fixed at 2$\theta$$_{out}$ $\sim$20$^{o}$  in the horizontal
direction. In this case the scattering vector {\emph{q}} lies
$\sim$10$^{o}$ to the film normal and, therefore, information is
obtained about the scattering planes almost parallel to the film
surface. ii) Grazing incidence in-plane (IP-XRD) geometry, where
both incident and diffracted beam make an angle of 0.05$^{o}$  in
vertical direction and detector was fixed at $\sim$20$^{o}$  in
the plane of the film.  In this case scattering vector lies almost
in the film plane and therefore, information is obtained about the
scattering planes perpendicular to the film surface, which was
kept horizontal.

The OP-XRD pattern of the F1 film exhibits three broad overlapping
peaks, which corresponds to (101), (002) and (100) planes of hcp
Co. In addition a faint peak around 3.56{\AA} is also visible
which corresponds to (200) reflection of fcc phase. Further, (111)
reflection of fcc phase at \emph{q}$~\sim$3.08{\AA} is very close
to (002) reflection of hcp phase at \emph{q}$~\sim$3.10{\AA} and
the two cannot be differentiated because of the large width of the
peaks. Thus the film F1 consist of mainly hcp phase with small
content of fcc phase.  After thermal annealing (film F2)
diffraction pattern gets changed significantly;  The peaks around
q=3.087{\AA} and \emph{q}=3.56 {\AA} become prominent, while the
intensity of the hcp (100) and hcp (101) peaks at
$\emph{q}$$~\sim$3.016 {\AA} and $\emph{q}$$~\sim$3.271 {\AA} get
reduced significantly. Thus thermal annealing results in
transformation of hcp phase of Co to fcc phase. This is in
agreement with earlier studies where thermal annealing above
350$^{o}$ C results in formation of fcc
phase.~\cite{DKumar:Co:Ann,JDAP09:Co-Ann} In film F3, the relative
intensities of hcp (100) and hcp (101) peaks at
$\emph{q}$$~\sim$3.016{\AA} and $\emph{q}$$~\sim$3.271 {\AA} again
shows an increase. Further the position of the peak at
$\emph{q}$$~\sim$3.087{\AA} shows a shift towards higher \emph{q},
suggesting that the peak now has a higher contribution of hcp
(002) reflection. Thus the Co film deposited on oxide layer is
predominantly in hcp phase. Figure 3(d) shows the IP- XRD of film
F3. A comparison of IP-XRD and OP-XRD of film F3 in table-I
clearly shows that the intensity (I$_{A}$) of hcp (002) peak is
substantially lower in the IP-XRD as compared to that in the
OP-XRD. This suggest that c-axis of top hcp-Co layer is
preferentially oriented out-of-plane.
\begin{table}
\caption{\label{tab:table1}. Results of fitting of XRD patterns
taken in in-plane (IP) and out of plane (OP) geometry for sample
F3; typical error bars, as obtained from the least square fitting
of the XRD data approx. $\pm$2$\%$  in the value of normalized
area, \emph{I$_{A}$}.}

\begin{ruledtabular}
\begin{tabular}{lcccc}
            &\multicolumn{1}{c}{q} &\multicolumn{1}{c}{\emph{I$_{A}$}} & \multicolumn{1}{c}{hkl } & \multicolumn{1}{c}{Co-phase}\\
    {Geometry} &   ({\AA}$^{-1}$) & {(\%)} \\

 \hline

              & 2.59     & 08    & 101        & hcp   \\
OP-XRD       & 3.11     & 64    & 111+002    & fcc+hcp   \\
              & 3.28     & 17    & 100        & hcp   \\
              & 3.58     & 11    & 200        & fcc   \\
 \hline
              & 2.96     & 04    & 101        & hcp   \\
IP-XRD       & 3.12     & 27    & 111+002    & fcc+hcp   \\
              & 3.30     & 49    & 100        & hcp  \\
              & 3.57     & 20    & 200        & fcc   \\

  \end{tabular}
  \end{ruledtabular}
  \end{table}

It may be noted that the hexagonal Co possesses an inherent
uniaxial magnetic anisotropy with c-axis being the easy
magnetization direction.~\cite{Sarath:APL} Therefore, in the
present case textured growth of top Co layer on CoO layer leads to
a volume contributed perpendicular magnetic anisotropy.The
possible cause of the textured growth of hcp-Co on oxide surface
can be as follows. During the initial stages of Co deposition
hybridization between Co and O at the interface leads to set Co-O
bonds perpendicular to the
interface.~\cite{PRB09:Co-Alox,PRB11:FE-Mgo,APE10:Qin-LiLv,nistor:APL9}
Perpendicular Co orbital moment at the interface would help to
grow hcp-Co with c-axis aligned preferentially perpendicular to
the film surface due to a significant lowering of the anisotropy
energy.~\cite{PRL98:CoPt} A similar effect has been predicted by
H. Ouyang \emph{et al.}~\cite{PRB10:NiO} through electronic
structure calculations in case of Co/Pt multilayer where interface
hybridization was tailored by preferred orientation of a NiO
buffer layer.

In conclusion, the growth behavior of Co thin film on CoO surface
has been studied \emph{in-situ} using MOKE. A clear perpendicular
magnetic anisotropy with easy axis perpendicular to the Co-CoO
interface has been observed in 104{\AA} thick Co film. This is in
contrast to the earlier studies, where PMA has been observed only
in ultra thin film of thickness $\sim$25{\AA} . Thus in present
case one is able to obtained PMA even in film thickness with an
order of magnitude higher. The origin of PMA in film is attributed
to the textured growth of hcp Co with c-axis perpendicular to the
film surface. Using depth resolved MOKE; it has been unambiguously
shown that only top Co layer possess PMA, whereas bottom layer was
found to be soft with in-plane magnetization without any influence
of top layer. It is important to mentioned that most of the
insulating layers used in TMR structures are metal oxide layers,
therefore it could be possible to deposit thick magnetic
electrode, which may lead to increased effective spin polarization
and hence TMR value.

%

\end{document}